\documentclass[amsmath,prb,superscriptaddress,preprint]{revtex4}

\usepackage{amsfonts}
\usepackage{amssymb}
\usepackage{graphicx}

\begin{document}

\title{Towards Noise Simulation in Interacting Nonequilibrium Systems Strongly Coupled to Baths}

\author{Kuniyuki Miwa}
\affiliation{Surface and Interface Science Laboratory, RIKEN, Wako, Saitama 351-0198, Japan}
\author{Feng Chen}
\affiliation{Department of Physics, University of California San Diego, La Jolla, CA 92093, USA}
\author{Michael Galperin}
\email{migalperin@ucsd.edu}
\affiliation{Department of Chemistry \& Biochemistry, University of California San Diego, La Jolla, CA 92093, USA}

\begin{abstract}
Progress in experimental techniques at nanoscale made measurements of noise 
in molecular junctions possible. These data are
important source of information not accessible through average flux measurements.
Emergence of optoelectronics, recently shown possibility of strong light-matter couplings, 
and developments in the field of quantum thermodynamics are making counting statistics measurements
of even higher importance.  
Theoretical methods for noise evaluation in first principles simulations 
can be roughly divided into approaches applicable in the case of weak intra-system interactions,
and those treating strong interactions for systems weakly coupled to baths.
We argue that due to structure of its diagrammatic expansion and the fact of utilizing
many-body states as a basis of its formulation recently introduced nonequilibrium Hubbard Green 
functions formulation is a relatively inexpensive method suitable for evaluation of noise characteristics
in first principles simulations over wide range of parameters.
We illustrate viability of the approach by simulations of noise and noise spectrum
within generic models for non-, weakly and strongly interacting systems.
Results of the simulations are compared to exact data (where available) and to simulations 
performed within approaches best suited for each of the three parameter regimes.
\end{abstract}

\maketitle

\section*{Introduction}
Progress in experimental techniques at nanoscale resulted in possibility to get information 
on transport characteristics of nanojunctions beyond average flux measurements.\cite{BlanterButtikerPR00}
In particular, shot noise measurements in single molecule
junctions were recently reported in the literature.\cite{RuitenbeekNL06} 
Shot noise (second cumulant in the full counting statistics of quantized charge transport) 
yields information not accessible through average flux measurements; 
it is also more sensitive to intra-molecular interactions such as, e.g.,
intra-system Coulomb repulsion,\cite{SchonenbergerPRL95} magnetism,\cite{TalRuitenbeekPRB13,BerndtPRL15,ArakawaPRL15}
and intra-molecular electron-vibration interactions.\cite{Tal2008,YeyatiRutenbeekPRL12}
Probing current-carrying molecular junctions by optical means, a new field of research coined
molecular optoelectronics,\cite{MGANPCCP12,MGChemSocRev17}
was demonstrated as another way to probe shot noise and charge fluctuations  
in nanojunctions,\cite{BerndtPRL10,PortierPRL10,BerndtPRL12}
and theory of light emission from quantum noise was recently formulated.\cite{KaasbjergNitzanPRL15}
Finally, recent experiments on strong\cite{SchwartzEbbesenAngChemIntEd12,SchwartzEbbesonChemPhysChem13,SchwartzEbbesenAdvMat13} and 
ultra-strong\cite{SchwartzEbbesenPRL11} light-molecules coupling in nano cavities
makes prospects of similar measurements in molecular junctions feasible.
Full counting statistics of both charge transport and photon flux (as well as their cross-correlations)
is especially important in this regime. 

The theoretical concept of full counting statistics was originally proposed by Levitov and 
Lesovik\cite{Levitov1993,Levitov1996}. and further developed in a numerous
studies.\cite{BlanterButtikerPR00} In particular, within the standard nonequilibrium Green function 
(NEGF) formulation the concept was considered in the work by Gogolin and 
Komnik\cite{GogolinKomnikPRB06}.  For noninteracting systems the formulation is exact,
and both steady-state\cite{GogolinKomnikPRB06,SchonhammerPRB07,SchonhammerJPCM09} and transient regime\cite{TangWangPRB14,TangWangPRB16} 
considerations are available in the literature. 
Accounting for intra-system interactions is a nontrivial task. Currently two main approaches are:
perturbation theory treatments for weak (relative to the system-baths coupling) 
interactions\cite{BoGalperinPRB97,MeirGolubPRL02,DiVentraPRL05,MGNitzanRatnerPRB06noise,SouzaJauhoEguesPRB08,NovotnyBelzigPRL09,KomnikPRB09,NovotnyBelzigPRB10,NovotnyBelzigPRB11}
and quantum master (or even rate) equations (QME) for strong interactions 
(and negligible system-baths coupling).\cite{DongPRL05,BelzigPRB05,vonOppenPRB06}
In the former and within the NEGF, self-consistent FCS formulations\cite{ParkMG_FCS_PRB11,LiEPJB13} 
(as required for an approximation to be conserving\cite{BaymKadanoffPR61,BaymPR62})  
were considered, and first {\em ab initio} simulations were performed.\cite{AvrillerFrederiksenPRB12}
The latter found numerous applications in developing field of quantum 
thermodynamics.\cite{EspositoRMP09}
Both approaches have their limitations: the former being formulated in the language of 
elementary excitations is inconvenient in treating strong intra-system interactions,
while the latter which utilizes system many-body states has difficulty in accounting 
for strong system-bath coupling. In this respect an important extension of the latter
approach is formulation of the real time perturbation theory (and similar approaches),\cite{SchonPRB94,SchonPRL96,SchoellerSchonPRB96,PedersenWackerPRB05,WackerPRB11,GrifoniEPJB13,GrifoniPRB15}
which allowed evaluation of shot noise  accounting for system-bath coupling within a systematic
perturbation expansion.\cite{SchonPRB03} 
 
Here we utilize recently proposed nonequilibrium diagrammatic technique for 
Hubbard Green functions (Hubbard NEGF)\cite{ChenOchoaMGJCP17} 
to simulate FCS and noise spectrum in model junctions.
We note in passing that earlier studies of Hubbard NEGF\cite{SandalovIJQC03,Fransson_2010} were 
restricted to evaluation of two-time  correlation functions only, 
while simulation of noise spectrum requires evaluation of multi-time
correlation functions. The latter is only possible within the diagrammatic consideration
(for detailed comparison between earlier studies and diagrammatic approach to Hubbard NEGF  
see Ref.~\cite{ChenOchoaMGJCP17}). 
Contrary to the standard NEGF, which considers multi-time correlation functions of 
elementary (de-)excitation  operators  $\hat d_i^\dagger$ ($\hat d_i$) - here $i$ is single electron orbital, 
the Hubbard NEGF deals with correlation functions of projection operators 
$\hat X_{S_1S_2}=\lvert S_1\rangle\langle S_2\rvert$ - here $\lvert S_{1,2}\rangle$ are many-body
states of the system. Similarity between the two approaches (illustrated by spectral
decomposition $\hat d_i=\sum_{S_1,S_2}\langle S_1\rvert\hat d_i\lvert S_2\rangle\,\Hat X_{S_1S_2}$)
indicates that structure of diagrammatic technique for the Hubbard NEGF should (to some extent)
resemble that of the standard NEGF (although expansions themselves are different:
Hubbard NEGF considers perturbation series in system-bath coupling,
while NEGF expands in small intra-system interaction). Similar way of
treating self-energies in the two techniques makes expectation that Hubbard NEGF
will be successful in accounting for relatively strong system-bath couplings plausible. 
At the same time formulation in the language of many-body states makes the Hubbard NEGF 
similar to  the QME approaches. Thus one may expect that accounting for strong intra-system 
interactions is feasible within the Hubbard NEGF.
Note that similar to the real time perturbation theory of 
Refs.~\cite{SchonPRB94,SchonPRL96,SchoellerSchonPRB96,SchonPRB03}
expansion is systematic.
Note also that the Hubbard NEGF approaches
in lowest orders of expansion is still simple enough to be applicable in realistic
simulations, thus paving a way to {\em ab initio} calculations of noise beyond weak interaction
limit of Ref.~\cite{AvrillerFrederiksenPRB12}.  

Below we present noise simulation in model systems within the Hubbard NEGF comparing 
results to those of other approaches. The presentation is followed by discussion of the approach
indicating its strong and weak sides and pointing to directions of further research.
At the end we briefly introduce the technique, details of the Hubbard NEGF and its application 
to noise simulation can be found in Ref.~\cite{ChenOchoaMGJCP17} and in 
the Supplementary Information.

\section*{Results}
We consider Anderson impurity as a junction model. Its Hamiltonian is 
\begin{equation}
\begin{split}
 \hat H=&\sum_{\sigma=\uparrow,\downarrow}\varepsilon_\sigma\hat n_\sigma
 + U\hat n_\uparrow\hat n_\downarrow
 +\sum_{\begin{subarray}{c} k\in L,R\\ \sigma=\uparrow,\downarrow\end{subarray}}\varepsilon_{k\sigma}\hat n_{k\sigma} +
 \sum_{\begin{subarray}{c} k\in L,R\\ \sigma_1,\sigma_2=\uparrow,\downarrow\end{subarray}}
 \bigg(V_{\sigma_1,k\sigma_2}\hat d^\dagger_{\sigma_1}\hat c_{k\sigma_2} + H.c.\bigg)
 \\
 \equiv& \sum_{S}E_S\hat X_{SS} + 
 \sum_{\begin{subarray}{c}k\in L,R; m\\ \sigma=\uparrow,\downarrow\end{subarray}}
 \varepsilon_{k\sigma}\hat n_{k\sigma} + 
  \sum_{\begin{subarray}{c}k\in L,R; m\\ \sigma_1,\sigma_2=\uparrow,\downarrow\end{subarray}}
 \bigg(V_{m(\sigma_1),k\sigma_2} \hat X_m^\dagger\hat c_{k\sigma_2} + H.c.\bigg)
 \end{split}
\end{equation}
First line utilizes second quantization, while second employs many-body states.
Here $\hat d^\dagger_{\sigma}$ and $\hat c^\dagger_{k\sigma}$ are creation operators
for electron with spin $\sigma$ in the system or contact state $k$, respectively;
$\hat n_\sigma=\hat d^\dagger_\sigma\hat d_\sigma$ and 
$\hat n_{k\sigma}=\hat c_{k\sigma}^\dagger\hat c_{k\sigma}$,
$\lvert S\rangle=\lvert 0\rangle, \lvert a\rangle, \lvert b\rangle, \lvert 2\rangle$
are empty, spin-up, spin-down, and double occupied many-body states 
with energies $E_0=0$, $E_a=\varepsilon_\uparrow$, $E_b=\varepsilon_\downarrow$,
$E_2=\varepsilon_\uparrow+\varepsilon_\downarrow+U$;
$m=0a, b2, 0b, a2$ are single electron transitions between pairs of many body states 
($0a$ and $b2$ are spin up transitions, $0b$ and $a2$ - spin down),
and $L$ and $R$ are left and right contacts. Within the model $U$ represents intra-system interaction,
system baths couplings are characterized by escape rate matrices,
$\Gamma^K_{\sigma_1\sigma_2}(E)=2\pi\sum_{k\in K;\sigma}V_{\sigma_1,k\sigma}V_{k\sigma,\sigma_2}\delta(E-\varepsilon_{k\sigma})$ ($K=L,R$), which are energy independent in the wide band approximation.

We simulate current and noise spectrum
at the left interface for steady-state transport situation
\begin{equation}
 I_L=\langle\hat I_L(0)\rangle
 \qquad 
 S_{LL}(\omega)=\frac{1}{2}\int_{-\infty}^{+\infty} dt\, e^{i\omega t} 
 \langle\{\delta\hat I_L(t);\delta\hat I_L(0)\}\rangle
 \qquad
 \hat I_L(t)=-\frac{d}{dt}\sum_{k\in L;\sigma}\hat n_{k\sigma}(t)
 \qquad
 \delta\hat I_L(t)=\hat I_L(t)-\langle\hat I_L(t)\rangle
\end{equation}
Green function techniques utilize exact (Meir-Wingreen) expression for the current, and approximate
(perturbative in system-baths couplings) derivation similar to that 
of Ref.~\cite{SouzaJauhoEguesPRB08} for noise spectrum.
In the Hubbard NEGF this derivation leads to diagrams presented in Fig.~\ref{fig1}d.
FCS is introduced as usual by dressing transfer matrix elements $V_{\sigma_1,k\sigma_2}$
($V_{m(\sigma_1),k\sigma_2}$) with contour branch dependent counting field $\lambda$
for $k\in L$. 
Zero frequency noise (second cumulant of the FCS) is obtained from derivative of the dressed current,
$S_{LL}(\omega=0)=-i\partial_\lambda I_L^\lambda\vert_{\lambda=0}$
(see, e.g., Ref.~\cite{EspMGPRB15} for details). 
 Simulations are performed  within the Hubbard NEGF, standard NEGF, and Lindblad/Redfield QME 
for non-interacting ($U=0$), weakly interacting ($U\leq\Gamma_0$), and
strongly interacting ($U\gg\Gamma_0$) cases (system-bath coupling strength $\Gamma_0$ is 
utilized as unit of energy).
Note that Lindblad/Redfield QME does not yield noise spectrum; current and zero-frequency noise
were simulated within the FCS (see, e.g., Ref.~\cite{EspositoMGJPCC10} for details). 
Results of the simulations are presented in units of $I_0\equiv e\Gamma_0/\hbar$ for current
and $S_0=e^2\Gamma_0/2\hbar$ for noise.

\subsection*{Non-interacting system}
We first present results of simulations for non-interacting case ($U=0$).
Note that all the standard NEGF results (including expression for noise spectrum) 
are exact in this case. 
We consider two sets of parameters representing degenerate and non-degenerate two-level systems.
Parameters for the degenerate model are (all energies are in units of $\Gamma_0$)
$\varepsilon_\uparrow=-5$, $\varepsilon_\downarrow=5$,
$\Gamma^K_{\uparrow\uparrow}=\Gamma^K_{\downarrow\downarrow}=1$
and $\Gamma^K_{\uparrow\downarrow}=\Gamma^K_{\downarrow\uparrow}=0.5$ ($K=L,R$).
Non-degenerate model is given by $\varepsilon_\uparrow=\varepsilon_\downarrow=-5$,
$\Gamma^K_{\uparrow\uparrow}=\Gamma^K_{\downarrow\downarrow}=1$ and
$\Gamma^K_{\uparrow\downarrow}=\Gamma^K_{\downarrow\uparrow}=0$.
In both models temperature was taken $k_bT=1/3$, Fermi energy chosen as origin $E_F=0$,
and bias applied symmetrically $\mu_{L,R}=E_F\pm |e|V_{sd}/2$.
Simulations were performed on energy grid spanning range from $-120$ to
$120$ with step $0.01$. Tolerance for convergence of 
the Hubbard NEGF scheme was $0.001$ for difference in density matrix values 
(state populations and coherences) in consequent steps of iteration. 
Counting field for numerical evaluation of FCS was chosen as $\lambda=0.01$.

Figure~\ref{fig2} presents results of FCS simulations for the degenerate model performed
within the Hubbard NEGF (dashed line) and standard NEGF (solid line). 
The Hubbard NEGF employs self-consistent FCS as described in, e.g., Ref.~\cite{ParkMG_FCS_PRB11}. 
FCS results were checked against analytical expressions known for noninteracting systems.
Standard NEGF (exact) results are reproduced very accurately by the approximate
Hubbard NEGF simulations for both current (Fig.~\ref{fig2}a) and zero-frequency
noise (Fig.~\ref{fig2}b). At resonant bias, $|e|V_{sd}=10\Gamma_0$ conductance
is maximum (inset in Fig.~\ref{fig2}a), while noise derivative has a dip (inset in Fig.~\ref{fig2}b),
as is expected from the Landauer expression for shot noise. 
Results for the non-degenerate model are similar.
Noise spectrum simulated for the two models within the Hubbard NEGF also reproduces 
exact (NEGF) results accurately (see Fig.~\ref{fig3}).

Note that another popular Green function methodology utilizing system many-body states,
the pseudoparticle NEGF (PP-NEGF), is widely employed as a standard
impurity solver in nonequilibrium dynamical mean field theory simulations.\cite{WernerRMP14}
Inability of the methodology to yield information on full counting statistics presumably
comes from its formulation within extended (unphysical) Hilbert space. 
However PP-NEGF can be utilized for simulations of current and noise spectrum.
In Supplementary Information we show that at the same (second order) level of diagrammatic 
perturbation expansion in system-baths couplings PP-NEGF reproduces current, 
but fails to reproduce noise spectrum. The reason is inability of the method to yield
correct spectral function even for non-interacting systems.

\subsection*{Weakly interacting system}
We now turn to weakly interacting case, where $U/\Gamma_0\ll1$ and perturbative expansion
in this small parameter within standard NEGF should yield reasonable results,\cite{StefanucciVanLeeuwen_2013} 
while Lindblad/Redfield QME is not expected to be accurate. 
Figure~\ref{fig4} shows results of FCS simulations for both models performed within 
the Hubbard NEGF (dashed line), standard NEGF (solid line), and Lindblad/Redfield QME (dotted line)
for a set of intra-system interaction strengths $U$.
Here both Green function techniques rely on self-consistency in numerical evaluation of FCS.
As expected results are quite similar for both Green function techniques and differ significantly
from the QME calculations. Naturally Hubbard and standard NEGF results coinciding at weak $U$
(see Figs.~\ref{fig4}a and \ref{fig4}e)
start to depart from each other at stronger interaction values. 
The difference is more pronounced 
in zero-frequency noise (main panels) than in the current (insets).
There is an interesting distinction between noise results for the two models:
comparing Figs.~\ref{fig4}c and \ref{fig4}g one sees that for $U=\Gamma_0$
the Hubbard NEGF noise result for degenerate model develops a dip at $|e|V_{sd}\sim 10\Gamma_0$, 
while no dip is observed in the non-degenerate case. 
The reason is difference in energetics of the two models: single electron transitions
$0a$, $b2$, $0b$, $a2$ for the non-degenerate model occur at 
$-5$, $-4$, $5$, $6$; while for degenerate model the transition are at 
$5$, $6$, $5$, $6$ (all energies are in units of $\Gamma_0$).
Thus by the time bias reaches resonant value of $|e|V_{sd}=10\Gamma_0$  
(when physical electronic population of the system becomes significant) 
channel $b2$ in non-degenerate case is open so that state $\lvert 2\rangle$
can be populated, so that no Coulomb blockade is observed. 
Degenerate model is Coulomb blockaded until $|e|V_{sd}=12\Gamma_0$.
Such suppression of shot noise due to charging effects was observed experimentally.\cite{SchonenbergerPRL95}
Note that the effect is seen already at $U=\Gamma_0/2$ (compare Figs.~\ref{fig4}b and \ref{fig4}f),
where perturbation expansion in $U/\Gamma_0$ should still be relatively accurate.
However standard NEGF does not reproduce the effect. The reason is a mean field
character of treatment of $U$ within second order diagrammatic expansion.     

\subsection*{Strong intra-system interaction}
Finally we consider strong interaction case, $U/\Gamma_0\gg 1$,
where Lindblad/Redfield is expected to be relatively accurate. 
Following Ref.~\cite{SouzaJauhoEguesPRB08} we consider a model 
with parameters (all energies are in units of $\Gamma_0$):
$\varepsilon_\uparrow=\varepsilon_\downarrow=50$, $U=100$,
$\Gamma^K_{\uparrow\uparrow}=\Gamma^K_{\downarrow\downarrow}=1$
and $\Gamma^K_{\uparrow\downarrow}=\Gamma^K_{\downarrow\uparrow}=0$ ($K=L,R$),
$k_BT=3$.
Figure~\ref{fig5}  (analog of Fig.~7 in Ref.~\cite{SouzaJauhoEguesPRB08})
shows current and zero frequency noise simulated
within the Hubbard NEGF (dashed line), standard NEGF (solid line),
and Lindblad/Redfield QME (dotted line). Also shown are analytical results of Ref.~\cite{SchonPRB03}.
Note that the standard NEGF consideration here follows Ref.~\cite{SouzaJauhoEguesPRB08}.
One sees that also here Hubbard NEGF follows quite closely the expected correct behavior 
(QME and analytical results). 

We note that diagrammatic Hubbard NEGF is a perturbative (in system-bath coupling) method,
and as such is not capable to treat strong system-bath correlations. In particular, the Kondo regime
is beyond capabilities of the method. 
At the same time there are many cases (including most experimental measurements of noise 
in molecular junctions) strong system-bath coupling is accompanied by non-negligible intra-system
interactions (e.g., electron-vibration coupling in the molecule) yet being outside of the Kondo regime.
Theoretical simulations of noise in such systems is complicated within both standard NEGF and QME
approaches. The Hubbard NEGF methodology presented yields a practical tool for first principle simulations in such systems.

\section*{Discussion}
Results of simulations show that the Hubbard NEGF is an inexpensive  method capable 
to reproduce satisfactory noise characteristics of junctions over a wide range of parameters. 
Indeed, one has to go only to lowest (second) order in perturbative expansion system-bath coupling 
to get reliable results. We think that this universality is due to basic components of 
the Hubbard NEGF formulation.
First, as a methodology utilizing many-body states of the system the Hubbard NEGF 
has a capability to account for strong intra-system interactions. This also makes it similar to 
QME considerations, which even at the lowest (second) order in system-bath coupling) 
are successfully utilized for FCS simulations in interacting systems weakly coupled to 
their surroundings. In particular, such QME methods are the cornerstone of quantum thermodynamics
and nonlinear optical spectroscopy considerations.  
At the same time as a Green functions formulation considering correlations of
single electron transitions Hubbard NEGF is close with the standard NEGF formulation.
The latter yields exact FCS results for  non-interacting systems, and was successfully applied 
in first principles simulations in the regime of strong coupling to baths and weak intra-system
interactions. These similarities with different techniques, each most suitable at opposite limits,
make the Hubbard NEGF a good candidate for a relatively inexpensive approach capable of predicting
noise properties in transport junctions over a wide range of parameters.   

It is interesting to note that in terms of structure of diagrammatic expansion
PP-NEGF (another popular many-body states Green function formulation), 
which considers correlation functions of many-body states, is closer to QME considerations.
Thus failure of the approach to predict zero-frequency noise in non-interacting systems
at the same (second) order of perturbation theory where Hubbard NEGF was successful,
is expected. Besides, the very formulation of the PP-NEGF in extended (unphysical) Hilbert space
presumably does not allow FCS formulation within the method. The latter was shown to work
well in the case of Hubbard NEGF.  

We show that the Hubbard NEGF yields accurate results over a broad range of parameters
already at the lowest (second) order in the system-bath coupling. 
For non-interacting case Hubbard NEGF yields results similar to the standard NEGF, which is exact in 
this limit. For weak intra-system interactions Hubbard NEGF is better than both the 
(same order of perturbation theory) standard NEGF and QME approaches.
At the same time for strong intra-system interactions real time perturbation theory works better than the present Hubbard NEGF consideration.
Thus careful (diagram to diagram) comparison of the two approaches is one of goals for
future studies. 

With respect to relevance to experiments we note that almost all noise measurements in 
molecular junctions are restricted to either non-interacting or relatively weak electron-vibration 
interactions\cite{RuitenbeekNL06,Tal2008,TalRuitenbeekPRL08,YeyatiRutenbeekPRL12} 
and are treated theoretically within second order perturbation theory.\cite{NovotnyBelzigPRL09,NovotnyBelzigPRB10,NovotnyBelzigPRB11,AvrillerFrederiksenPRB12}
Hubbard NEGF is a scheme applicable in such experimentally relevant situations
capable to go beyond lowest order perturbative schemes in intra-system interactions. 
It also may serve as a useful theoretical tool for investigation of (soon expected) 
measurements of strong light-matter  interactions\cite{SchwartzEbbesenAngChemIntEd12,SchwartzEbbesonChemPhysChem13,SchwartzEbbesenAdvMat13,SchwartzEbbesenPRL11} in molecular junctions. 

\section*{Methods}
Details on standard NEGF simulations can be found in, e.g., Refs.~\cite{SouzaJauhoEguesPRB08,ParkMG_FCS_PRB11}.
Lindblad/Redfield simulations are discussed in, e.g., Refs.~\cite{EspositoRMP09,EspositoMGJPCC10}.
PP-NEGF formulation is discussed in, e.g., Refs.~\cite{WernerRMP14,WhiteGalperinPCCP12}.
Expressions for noise spectrum within the PP-NEGF can be found in Supporting Information. 
Here we briefly discuss the Hubbard NEGF approach.

\begin{figure}[ht]
\centering
\includegraphics[width=\linewidth]{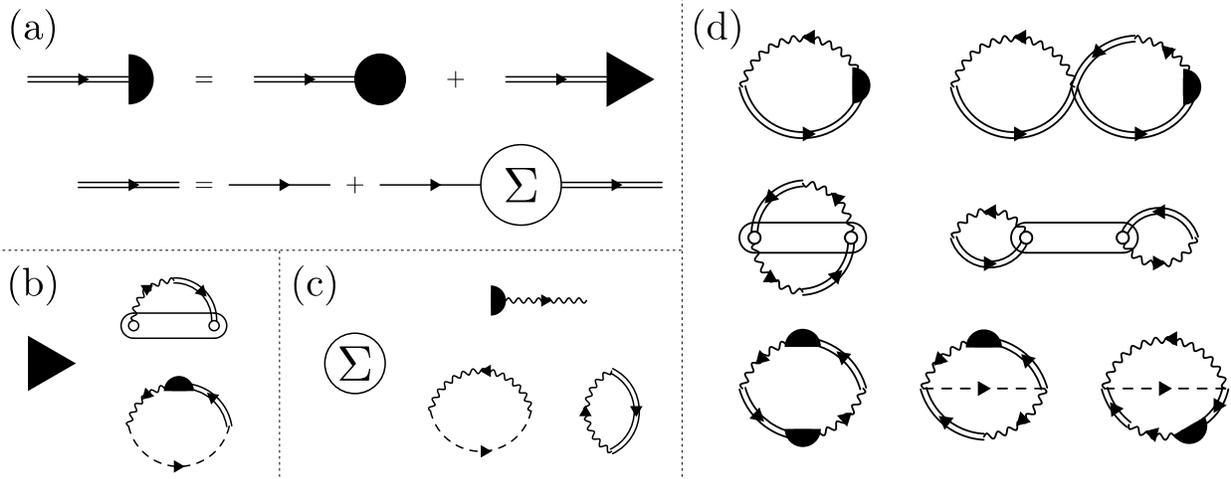}
\caption{
Hubbard NEGF. Shown are (a) generalized Dyson equation and
second order dressed diagrams for  (b) vertex $\Delta$ (triangle) and (c) self-energy $\Sigma$.
Panel (d) presents diagrams utilized in noise spectrum simulations.
Directed single (double) solid line represents zero-order (dressed) locator $g^{(0)}$ ($g$),
directed dashed line stands for two-electron Green function, and directed wavy line indicates
electron-electron correlation function yielding influence of contacts (in standard NEGF this
function is called self-energy due to coupling to contacts).
Circle represents spectral weight $F$, semi-circle stands for the strength operator $P=F+\Delta$,
and oval  is intra-molecular correlation function.
See Methods and Supporting Information for details.
}
\label{fig1}
\end{figure}

\begin{figure}[ht]
\centering
\includegraphics[width=0.5\linewidth]{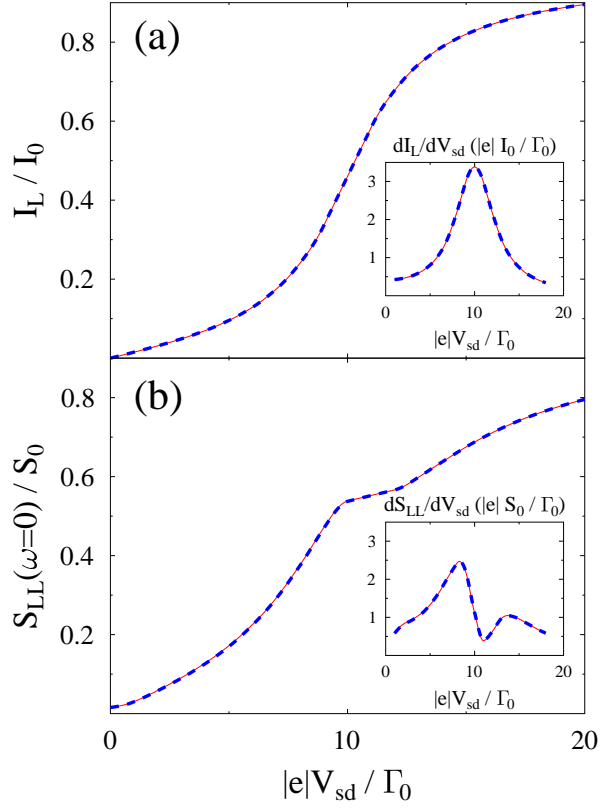}
\caption{
Degenerate non-interacting two-level system.
Shown are (a) current and (b) zero frequency noise obtained employing
full counting statistics within the Hubbard NEGF (dashed line, blue) 
and standard NEGF (solid line, red) approaches. For non-interacting case the latter is exact.
Insets show conductance and differential noise, respectively. See text for parameters.
}
\label{fig2}
\end{figure}

\begin{figure}[ht]
\centering
\includegraphics[width=\linewidth]{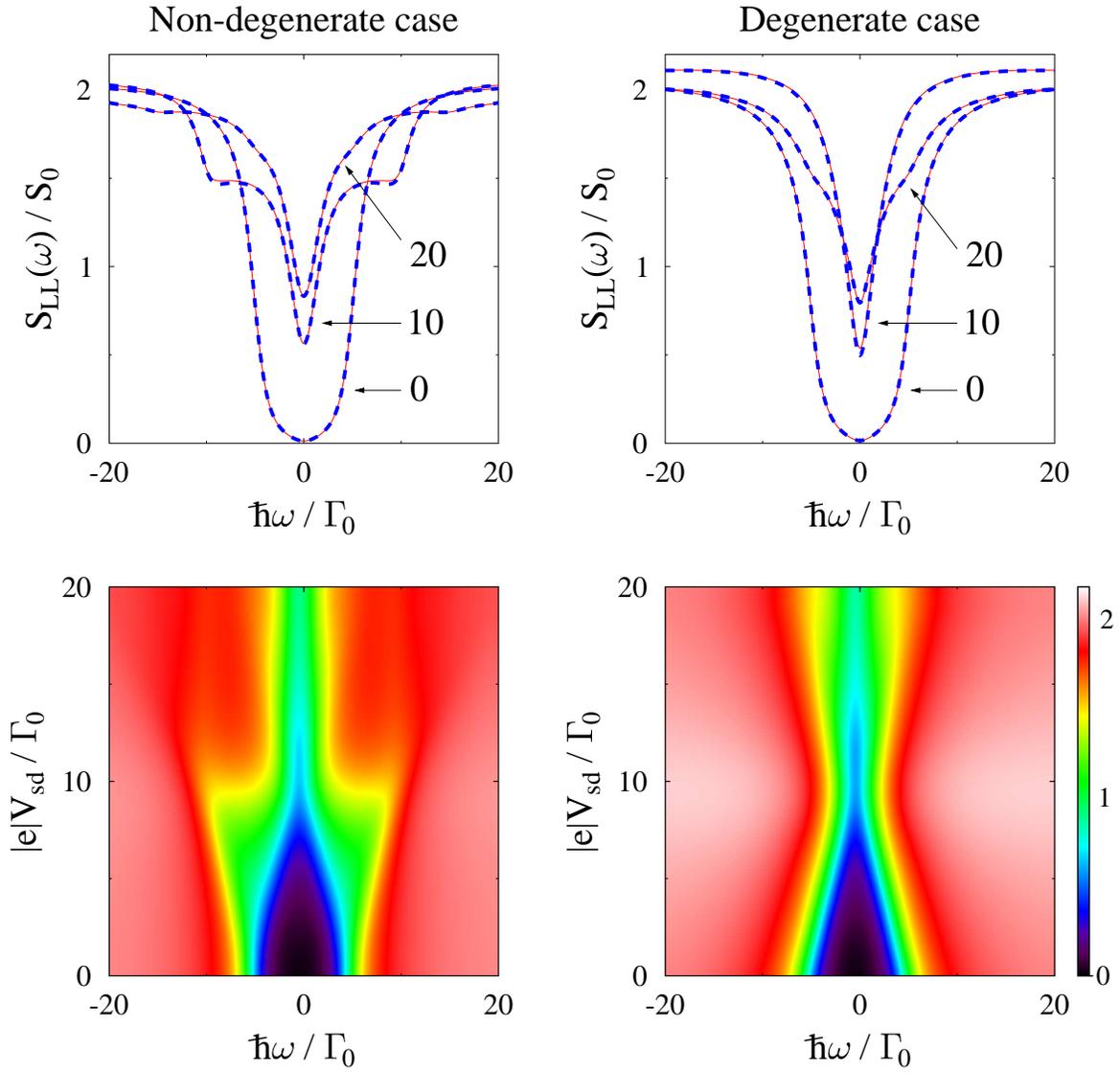}
\caption{
Non-degenerate (left) and degenerate (right) non-interacting two-level systems. 
Shown is noise spectrum simulated within the Hubbard NEGF utilizing diagrams in Fig.~\ref{fig1}d.
Top panels (horizontal cuts of lower maps) compare the Hubbard NEGF (dashed line, blue) 
with exact results (solid line, red) for three biases. See text for parameters. 
}
\label{fig3}
\end{figure}

\begin{figure}[ht]
\centering
\includegraphics[width=\linewidth]{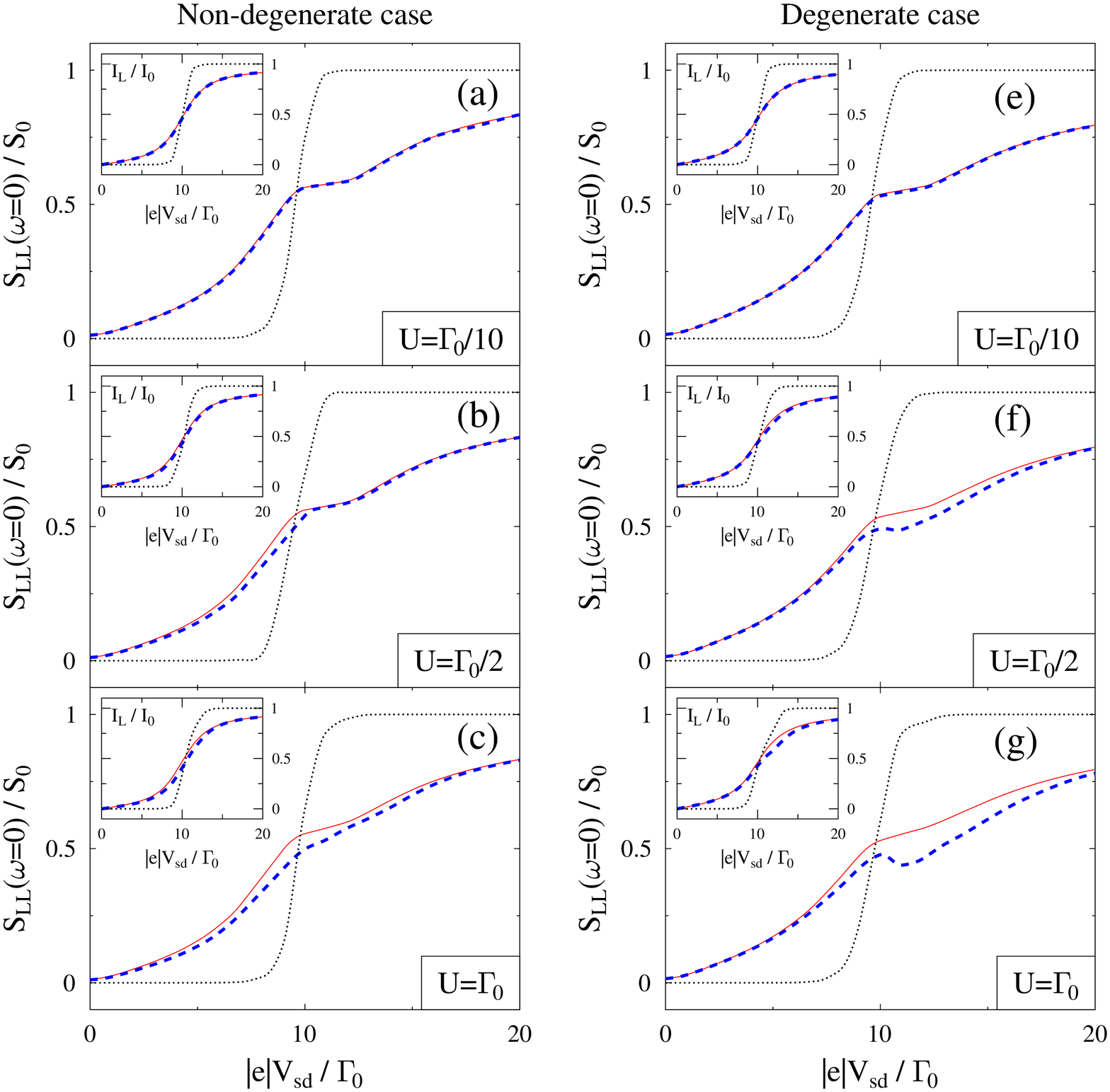}
\caption{
Non-degenerate (a-c) and degenerate (e-g) Hubbard model in the weak interaction $U\leq\Gamma_0$
regime. Shown are current (insets) and zero-frequency noise vs. applied bias.
Simulations are performed utilizing the full counting statistics within 
the Hubbard NEGF (dashed line, blue), standard NEGF (solid line, red), and 
Lindblad/Redfield QME (dotted line, black). See text for parameters.
}
\label{fig4}
\end{figure}

\begin{figure}[ht]
\centering
\includegraphics[width=0.5\linewidth]{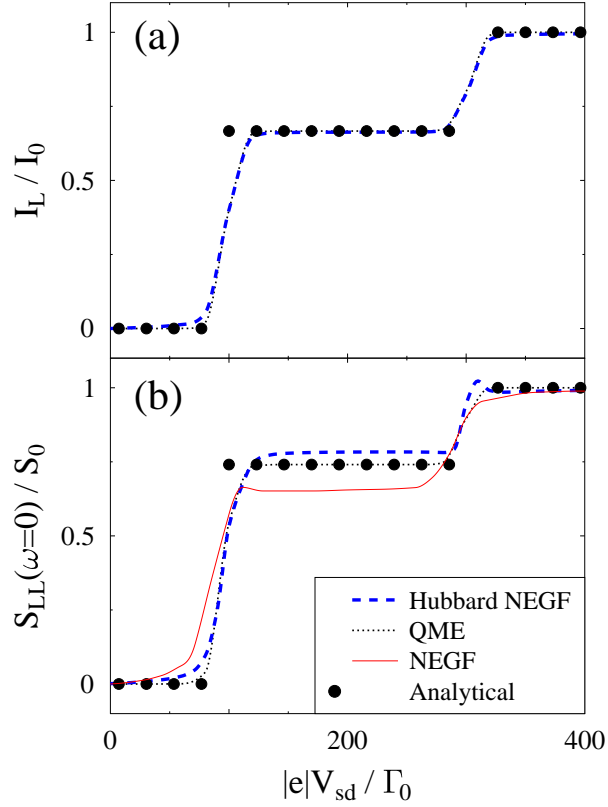}
\caption{
Hubbard model in the strong interaction regime $U\gg\Gamma_0$. 
Shown are (a) current and (b) zero-frequency noise vs. applied bias.
calculated within the Hubbard NEGF (dashed line, blue), standard NEGF (solid line, red),
and Lindblad/Redfield QME (dotted line, black). 
Solid circles (black) indicate analytical result as derived in Ref.~\cite{SchonPRB03}  
within the lowest (first-) order perturbation theory in $\Gamma_0$.
Noise simulations are performed within the Hubbard NEGF
and standard NEGF utilizing noise spectrum diagrams  of Fig.~\ref{fig1}d and Ref.~\cite{SouzaJauhoEguesPRB08}, respectively;
simulations within Lindblad/Redfield QME utilize full counting statistics.
See text for parameters.
}
\label{fig5}
\end{figure}

Hubbard NEGF considers correlation function of Hubbard operators defined on the Keldysh contour as
\begin{equation}
 G_{m_1m_2}(\tau_1,\tau_2) = -i\langle T_c\,\hat X_{m_1}(\tau_1)\,\hat X_{m_2}^\dagger(\tau_2)\rangle
\end{equation} 
where $\tau_{1,2}$ are contour variables, $T_c$ is contour ordering operator,
$m_{1,2}$ are single electron transitions between many body states,
and $\hat X_{m}=\lvert S_{1}\rangle\langle S_2\rvert$ with $\lvert S_1\rangle$ containing
one electron less than $\lvert S_2\rangle$.
The Green function is obtained by solving a modified version of the Dyson type equation
(see Fig.~\ref{fig1}a)
\begin{equation}
\label{HubDyson}
\begin{split}
& G_{m_1m_2}(\tau_1,\tau_2) = \sum_{m_3}\int_c d\tau_3\, g_{m_1m_3}(\tau_1,\tau_3)\, P_{m_3m_2}(\tau_3,\tau_2)
\\
& g_{m_1m_2}(\tau_1,\tau_2) = g^{0}_{m_1m_2}(\tau_1,\tau_2) 
+ \sum_{m_3,m_4}\int_c d\tau_3\int_c d\tau_4\, 
g^{0}_{m_1m_3}(\tau_1,\tau_3)\,\Sigma_{m_3m_4}(\tau_3,\tau_4)\,g_{m_4m_2}(\tau_4,\tau_2)
\end{split}
\end{equation}
Here $g_{m_1m_2}(\tau_1,\tau_2)$ is the locator, $g^{0}_{m_1m_2}(\tau_1,\tau_2)$ is the locator
in the absence of coupling to the baths, $\Sigma_{m_1m_2}(\tau_1,\tau_2)$ is self-energy,
and $P_{m_1m_2}(\tau_1,\tau_2)$ is the strength operator.
The latter is sum of spectral weight $F_{m_1m_2}(\tau)$ and vertex $\Delta_{m_1m_2}(\tau_1,\tau_2)$
functions (dressed second order diagrams for vertex $\Delta$ and self-energy $\Sigma$ are 
shown in Figs.~\ref{fig1}b and \ref{fig1}c, respectively).  
Self-consistency of the approach comes form the fact that
both self-energy and strength operator depend on the Hubbard Green function.
After convergence Hubbard Green function is utilized in simulation of current and noise.
For FCS the Eqs.~(\ref{HubDyson}) are dressed with counting fields.
Details of the self-consistent procedure are given in Ref.~\cite{ChenOchoaMGJCP17}. 
Explicit expressions for noise spectrum in terms of Hubbard Green functions
are given in the Supporting Information. Fig.~\ref{fig1}d shows corresponding diagrams.

{\bf Data Availability Statement:} All data generated or analyzed during this study are included in this published article (and its Supplementary Information files).


\begin{thebibliography}{10}
\expandafter\ifx\csname url\endcsname\relax
  \def\url#1{\texttt{#1}}\fi
\expandafter\ifx\csname urlprefix\endcsname\relax\def\urlprefix{URL }\fi
\expandafter\ifx\csname doiprefix\endcsname\relax\def\doiprefix{DOI }\fi
\providecommand{\bibinfo}[2]{#2}
\providecommand{\eprint}[2][]{\url{#2}}

\bibitem{BlanterButtikerPR00}
\bibinfo{author}{Blanter, Y.~M.} \& \bibinfo{author}{Buttiker, M.}
\newblock \bibinfo{title}{Shot noise in mesoscopic conductors}.
\newblock \emph{\bibinfo{journal}{Phys. Rep.}} \textbf{\bibinfo{volume}{336}},
  \bibinfo{pages}{1--166} (\bibinfo{year}{2000}).

\bibitem{RuitenbeekNL06}
\bibinfo{author}{Djukic, D.} \& \bibinfo{author}{van Ruitenbeek, J.~M.}
\newblock \bibinfo{title}{Shot noise measurements on a single molecule}.
\newblock \emph{\bibinfo{journal}{Nano Lett.}} \textbf{\bibinfo{volume}{6}},
  \bibinfo{pages}{789--793} (\bibinfo{year}{2006}).

\bibitem{SchonenbergerPRL95}
\bibinfo{author}{Birk, H.}, \bibinfo{author}{de~Jong, M. J.~M.} \&
  \bibinfo{author}{Sch\"onenberger, C.}
\newblock \bibinfo{title}{Shot-noise suppression in the single-electron
  tunneling regime}.
\newblock \emph{\bibinfo{journal}{Phys. Rev. Lett.}}
  \textbf{\bibinfo{volume}{75}}, \bibinfo{pages}{1610--1613}
  (\bibinfo{year}{1995}).

\bibitem{TalRuitenbeekPRB13}
\bibinfo{author}{Kumar, M.} \emph{et~al.}
\newblock \bibinfo{title}{Shot noise and magnetism of {P}t atomic chains:
  Accumulation of points at the boundary}.
\newblock \emph{\bibinfo{journal}{Phys. Rev. B}} \textbf{\bibinfo{volume}{88}},
  \bibinfo{pages}{245431} (\bibinfo{year}{2013}).

\bibitem{BerndtPRL15}
\bibinfo{author}{Burtzlaff, A.}, \bibinfo{author}{Weismann, A.},
  \bibinfo{author}{Brandbyge, M.} \& \bibinfo{author}{Berndt, R.}
\newblock \bibinfo{title}{Shot noise as a probe of spin-polarized transport
  through single atoms}.
\newblock \emph{\bibinfo{journal}{Phys. Rev. Lett.}}
  \textbf{\bibinfo{volume}{114}}, \bibinfo{pages}{016602}
  (\bibinfo{year}{2015}).

\bibitem{ArakawaPRL15}
\bibinfo{author}{Arakawa, T.} \emph{et~al.}
\newblock \bibinfo{title}{Shot noise induced by nonequilibrium spin
  accumulation}.
\newblock \emph{\bibinfo{journal}{Phys. Rev. Lett.}}
  \textbf{\bibinfo{volume}{114}}, \bibinfo{pages}{016601}
  (\bibinfo{year}{2015}).

\bibitem{Tal2008}
\bibinfo{author}{Tal, O.}, \bibinfo{author}{Krieger, M.},
  \bibinfo{author}{Leerink, B.} \& \bibinfo{author}{van Ruitenbeek, J.~M.}
\newblock \bibinfo{title}{Electron-vibration interaction in single-molecule
  junctions: From contact to tunneling regimes}.
\newblock \emph{\bibinfo{journal}{Phys. Rev. Lett.}}
  \textbf{\bibinfo{volume}{100}}, \bibinfo{pages}{196804}
  (\bibinfo{year}{2008}).

\bibitem{YeyatiRutenbeekPRL12}
\bibinfo{author}{Kumar, M.}, \bibinfo{author}{Avriller, R.},
  \bibinfo{author}{Yeyati, A.~L.} \& \bibinfo{author}{van Ruitenbeek, J.~M.}
\newblock \bibinfo{title}{Detection of vibration-mode scattering in electronic
  shot noise}.
\newblock \emph{\bibinfo{journal}{Phys. Rev. Lett.}}
  \textbf{\bibinfo{volume}{108}}, \bibinfo{pages}{146602}
  (\bibinfo{year}{2012}).

\bibitem{MGANPCCP12}
\bibinfo{author}{Galperin, M.} \& \bibinfo{author}{Nitzan, A.}
\newblock \bibinfo{title}{Molecular optoelectronics: The interaction of
  molecular conduction junctions with light}.
\newblock \emph{\bibinfo{journal}{Phys. Chem. Chem. Phys.}}
  \textbf{\bibinfo{volume}{14}}, \bibinfo{pages}{9421--9438}
  (\bibinfo{year}{2012}).

\bibitem{MGChemSocRev17}
\bibinfo{author}{Galperin, M.}
\newblock \bibinfo{title}{Photonics and spectroscopy in nanojunctions: a
  theoretical insight}.
\newblock \emph{\bibinfo{journal}{Chem. Soc. Rev.}} \bibinfo{pages}{--}
  (\bibinfo{year}{2017}).

\bibitem{BerndtPRL10}
\bibinfo{author}{Schneider, N.~L.}, \bibinfo{author}{Schull, G.} \&
  \bibinfo{author}{Berndt, R.}
\newblock \bibinfo{title}{Optical probe of quantum shot-noise reduction at a
  single-atom contact}.
\newblock \emph{\bibinfo{journal}{Phys. Rev. Lett.}}
  \textbf{\bibinfo{volume}{105}}, \bibinfo{pages}{026601}
  (\bibinfo{year}{2010}).

\bibitem{PortierPRL10}
\bibinfo{author}{Zakka-Bajjani, E.} \emph{et~al.}
\newblock \bibinfo{title}{Experimental determination of the statistics of
  photons emitted by a tunnel junction}.
\newblock \emph{\bibinfo{journal}{Phys. Rev. Lett.}}
  \textbf{\bibinfo{volume}{104}}, \bibinfo{pages}{206802}
  (\bibinfo{year}{2010}).

\bibitem{BerndtPRL12}
\bibinfo{author}{Schneider, N.~L.}, \bibinfo{author}{L\"u, J.~T.},
  \bibinfo{author}{Brandbyge, M.} \& \bibinfo{author}{Berndt, R.}
\newblock \bibinfo{title}{Light emission probing quantum shot noise and charge
  fluctuations at a biased molecular junction}.
\newblock \emph{\bibinfo{journal}{Phys. Rev. Lett.}}
  \textbf{\bibinfo{volume}{109}}, \bibinfo{pages}{186601}
  (\bibinfo{year}{2012}).

\bibitem{KaasbjergNitzanPRL15}
\bibinfo{author}{Kaasbjerg, K.} \& \bibinfo{author}{Nitzan, A.}
\newblock \bibinfo{title}{Theory of light emission from quantum noise in
  plasmonic contacts: Above-threshold emission from higher-order
  electron-plasmon scattering}.
\newblock \emph{\bibinfo{journal}{Phys. Rev. Lett.}}
  \textbf{\bibinfo{volume}{114}}, \bibinfo{pages}{126803}
  (\bibinfo{year}{2015}).

\bibitem{SchwartzEbbesenAngChemIntEd12}
\bibinfo{author}{Hutchison, J.~A.}, \bibinfo{author}{Schwartz, T.},
  \bibinfo{author}{Genet, C.}, \bibinfo{author}{Devaux, E.} \&
  \bibinfo{author}{Ebbesen, T.~W.}
\newblock \bibinfo{title}{Modifying chemical landscapes by coupling to vacuum
  fields}.
\newblock \emph{\bibinfo{journal}{Angew. Chem. Int. Ed.}}
  \textbf{\bibinfo{volume}{51}}, \bibinfo{pages}{1592--1596}
  (\bibinfo{year}{2012}).

\bibitem{SchwartzEbbesonChemPhysChem13}
\bibinfo{author}{Schwartz, T.} \emph{et~al.}
\newblock \bibinfo{title}{Polariton dynamics under strong light–molecule
  coupling}.
\newblock \emph{\bibinfo{journal}{ChemPhysChem}} \textbf{\bibinfo{volume}{14}},
  \bibinfo{pages}{125--131} (\bibinfo{year}{2013}).

\bibitem{SchwartzEbbesenAdvMat13}
\bibinfo{author}{Hutchison, J.~A.} \emph{et~al.}
\newblock \bibinfo{title}{Tuning the work-function via strong coupling}.
\newblock \emph{\bibinfo{journal}{Adv. Mater.}} \textbf{\bibinfo{volume}{25}},
  \bibinfo{pages}{2481--2485} (\bibinfo{year}{2013}).

\bibitem{SchwartzEbbesenPRL11}
\bibinfo{author}{Schwartz, T.}, \bibinfo{author}{Hutchison, J.~A.},
  \bibinfo{author}{Genet, C.} \& \bibinfo{author}{Ebbesen, T.~W.}
\newblock \bibinfo{title}{Reversible switching of ultrastrong light-molecule
  coupling}.
\newblock \emph{\bibinfo{journal}{Phys. Rev. Lett.}}
  \textbf{\bibinfo{volume}{106}}, \bibinfo{pages}{196405}
  (\bibinfo{year}{2011}).

\bibitem{Levitov1993}
\bibinfo{author}{Levitov, L.~S.} \& \bibinfo{author}{Lesovik, G.~B.}
\newblock \bibinfo{title}{Charge distribution in quantum shot noise}.
\newblock \emph{\bibinfo{journal}{JETP Lett.}} \textbf{\bibinfo{volume}{58}},
  \bibinfo{pages}{230--235} (\bibinfo{year}{1993}).

\bibitem{Levitov1996}
\bibinfo{author}{Levitov, L.~S.}, \bibinfo{author}{Lee, H.} \&
  \bibinfo{author}{Lesovik, G.~B.}
\newblock \bibinfo{title}{Electron counting statistics and coherent states of
  electric current}.
\newblock \emph{\bibinfo{journal}{JMP}} \textbf{\bibinfo{volume}{37}},
  \bibinfo{pages}{4845--4866} (\bibinfo{year}{1996}).

\bibitem{GogolinKomnikPRB06}
\bibinfo{author}{Gogolin, A.~O.} \& \bibinfo{author}{Komnik, A.}
\newblock \bibinfo{title}{Towards full counting statistics for the {A}nderson
  impurity model}.
\newblock \emph{\bibinfo{journal}{Phys. Rev. B}} \textbf{\bibinfo{volume}{73}},
  \bibinfo{pages}{195301} (\bibinfo{year}{2006}).

\bibitem{SchonhammerPRB07}
\bibinfo{author}{Sch\"onhammer, K.}
\newblock \bibinfo{title}{Full counting statistics for noninteracting fermions:
  Exact results and the {L}evitov-{L}esovik formula}.
\newblock \emph{\bibinfo{journal}{Phys. Rev. B}} \textbf{\bibinfo{volume}{75}},
  \bibinfo{pages}{205329} (\bibinfo{year}{2007}).

\bibitem{SchonhammerJPCM09}
\bibinfo{author}{Sch\"onhammer, K.}
\newblock \bibinfo{title}{Full counting statistics for noninteracting fermions:
  exact finite-temperature results and generalized long-time approximation}.
\newblock \emph{\bibinfo{journal}{J. Phys.: Condens. Matter}}
  \textbf{\bibinfo{volume}{21}}, \bibinfo{pages}{495306}
  (\bibinfo{year}{2009}).

\bibitem{TangWangPRB14}
\bibinfo{author}{Tang, G.-M.} \& \bibinfo{author}{Wang, J.}
\newblock \bibinfo{title}{Full-counting statistics of charge and spin transport
  in the transient regime: A nonequilibrium {G}reen's function approach}.
\newblock \emph{\bibinfo{journal}{Phys. Rev. B}} \textbf{\bibinfo{volume}{90}},
  \bibinfo{pages}{195422} (\bibinfo{year}{2014}).

\bibitem{TangWangPRB16}
\bibinfo{author}{Yu, Z.}, \bibinfo{author}{Tang, G.-M.} \&
  \bibinfo{author}{Wang, J.}
\newblock \bibinfo{title}{Full-counting statistics of transient energy current
  in mesoscopic systems}.
\newblock \emph{\bibinfo{journal}{Phys. Rev. B}} \textbf{\bibinfo{volume}{93}},
  \bibinfo{pages}{195419} (\bibinfo{year}{2016}).

\bibitem{BoGalperinPRB97}
\bibinfo{author}{Bo, O.~L.} \& \bibinfo{author}{Galperin, Y.}
\newblock \bibinfo{title}{Low-frequency shot noise in phonon-assisted resonant
  magnetotunneling}.
\newblock \emph{\bibinfo{journal}{Phys. Rev. B}} \textbf{\bibinfo{volume}{55}},
  \bibinfo{pages}{1696--1706} (\bibinfo{year}{1997}).

\bibitem{MeirGolubPRL02}
\bibinfo{author}{Meir, Y.} \& \bibinfo{author}{Golub, A.}
\newblock \bibinfo{title}{Shot noise through a quantum dot in the {K}ondo
  regime}.
\newblock \emph{\bibinfo{journal}{Phys. Rev. Lett.}}
  \textbf{\bibinfo{volume}{88}}, \bibinfo{pages}{116802}
  (\bibinfo{year}{2002}).

\bibitem{DiVentraPRL05}
\bibinfo{author}{Chen, Y.-C.} \& \bibinfo{author}{Di~Ventra, M.}
\newblock \bibinfo{title}{Effect of electron-phonon scattering on shot noise in
  nanoscale junctions}.
\newblock \emph{\bibinfo{journal}{Phys. Rev. Lett.}}
  \textbf{\bibinfo{volume}{95}}, \bibinfo{pages}{166802}
  (\bibinfo{year}{2005}).

\bibitem{MGNitzanRatnerPRB06noise}
\bibinfo{author}{Galperin, M.}, \bibinfo{author}{Nitzan, A.} \&
  \bibinfo{author}{Ratner, M.~A.}
\newblock \bibinfo{title}{Inelastic tunneling effects on noise properties of
  molecular junctions}.
\newblock \emph{\bibinfo{journal}{Phys. Rev. B}} \textbf{\bibinfo{volume}{74}},
  \bibinfo{pages}{075326} (\bibinfo{year}{2006}).

\bibitem{SouzaJauhoEguesPRB08}
\bibinfo{author}{Souza, F.~M.}, \bibinfo{author}{Jauho, A.~P.} \&
  \bibinfo{author}{Egues, J.~C.}
\newblock \bibinfo{title}{Spin-polarized current and shot noise in the presence
  of spin flip in a quantum dot via nonequilibrium {G}reen's functions}.
\newblock \emph{\bibinfo{journal}{Phys. Rev. B}} \textbf{\bibinfo{volume}{78}},
  \bibinfo{pages}{155303} (\bibinfo{year}{2008}).

\bibitem{NovotnyBelzigPRL09}
\bibinfo{author}{Haupt, F.}, \bibinfo{author}{Novotn{\' y}, T.} \&
  \bibinfo{author}{Belzig, W.}
\newblock \bibinfo{title}{Phonon-assisted current noise in molecular
  junctions}.
\newblock \emph{\bibinfo{journal}{Phys. Rev. Lett.}}
  \textbf{\bibinfo{volume}{103}}, \bibinfo{pages}{136601}
  (\bibinfo{year}{2009}).

\bibitem{KomnikPRB09}
\bibinfo{author}{Schmidt, T.~L.} \& \bibinfo{author}{Komnik, A.}
\newblock \bibinfo{title}{Charge transfer statistics of a molecular quantum dot
  with a vibrational degree of freedom}.
\newblock \emph{\bibinfo{journal}{Phys. Rev. B}} \textbf{\bibinfo{volume}{80}},
  \bibinfo{pages}{041307(R)} (\bibinfo{year}{2009}).

\bibitem{NovotnyBelzigPRB10}
\bibinfo{author}{Haupt, F.}, \bibinfo{author}{Novotn\'y, T. c.~v.} \&
  \bibinfo{author}{Belzig, W.}
\newblock \bibinfo{title}{Current noise in molecular junctions: Effects of the
  electron-phonon interaction}.
\newblock \emph{\bibinfo{journal}{Phys. Rev. B}} \textbf{\bibinfo{volume}{82}},
  \bibinfo{pages}{165441} (\bibinfo{year}{2010}).

\bibitem{NovotnyBelzigPRB11}
\bibinfo{author}{Novotn\'y, T. c.~v.}, \bibinfo{author}{Haupt, F.} \&
  \bibinfo{author}{Belzig, W.}
\newblock \bibinfo{title}{Nonequilibrium phonon backaction on the current noise
  in atomic-sized junctions}.
\newblock \emph{\bibinfo{journal}{Phys. Rev. B}} \textbf{\bibinfo{volume}{84}},
  \bibinfo{pages}{113107} (\bibinfo{year}{2011}).

\bibitem{DongPRL05}
\bibinfo{author}{Dong, B.}, \bibinfo{author}{Cui, H.~L.} \&
  \bibinfo{author}{Lei, X.~L.}
\newblock \bibinfo{title}{Pumped spin-current and shot-noise spectra of a
  single quantum dot}.
\newblock \emph{\bibinfo{journal}{Phys. Rev. Lett.}}
  \textbf{\bibinfo{volume}{94}}, \bibinfo{pages}{066601}
  (\bibinfo{year}{2005}).

\bibitem{BelzigPRB05}
\bibinfo{author}{Belzig, W.}
\newblock \bibinfo{title}{Full counting statistics of super-{P}oissonian shot
  noise in multilevel quantum dots}.
\newblock \emph{\bibinfo{journal}{Phys. Rev. B}} \textbf{\bibinfo{volume}{71}},
  \bibinfo{pages}{161301} (\bibinfo{year}{2005}).

\bibitem{vonOppenPRB06}
\bibinfo{author}{Koch, J.}, \bibinfo{author}{von Oppen, F.} \&
  \bibinfo{author}{Andreev, A.~V.}
\newblock \bibinfo{title}{Theory of the {F}ranck-{C}ondon blockade regime}.
\newblock \emph{\bibinfo{journal}{Phys. Rev. B}} \textbf{\bibinfo{volume}{74}},
  \bibinfo{pages}{205438} (\bibinfo{year}{2006}).

\bibitem{ParkMG_FCS_PRB11}
\bibinfo{author}{Park, T.-H.} \& \bibinfo{author}{Galperin, M.}
\newblock \bibinfo{title}{Self-consistent full counting statistics of inelastic
  transport}.
\newblock \emph{\bibinfo{journal}{Phys. Rev. B}} \textbf{\bibinfo{volume}{84}},
  \bibinfo{pages}{205450} (\bibinfo{year}{2011}).

\bibitem{LiEPJB13}
\bibinfo{author}{Li, H.}, \bibinfo{author}{Agarwalla, B.~K.},
  \bibinfo{author}{Li, B.} \& \bibinfo{author}{Wang, J.-S.}
\newblock \bibinfo{title}{Cumulants of heat transfer across nonlinear quantum
  systems}.
\newblock \emph{\bibinfo{journal}{Eur. Phys. J. B}}
  \textbf{\bibinfo{volume}{86}}, \bibinfo{pages}{500} (\bibinfo{year}{2013}).

\bibitem{BaymKadanoffPR61}
\bibinfo{author}{Baym, G.} \& \bibinfo{author}{Kadanoff, L.~P.}
\newblock \bibinfo{title}{Conservation laws and correlation functions}.
\newblock \emph{\bibinfo{journal}{Phys. Rev.}} \textbf{\bibinfo{volume}{124}},
  \bibinfo{pages}{287--299} (\bibinfo{year}{1961}).

\bibitem{BaymPR62}
\bibinfo{author}{Baym, G.}
\newblock \bibinfo{title}{Self-consistent approximations in many-body systems}.
\newblock \emph{\bibinfo{journal}{Phys. Rev.}} \textbf{\bibinfo{volume}{127}},
  \bibinfo{pages}{1391--1401} (\bibinfo{year}{1962}).

\bibitem{AvrillerFrederiksenPRB12}
\bibinfo{author}{Avriller, R.} \& \bibinfo{author}{Frederiksen, T.}
\newblock \bibinfo{title}{Inelastic shot noise characteristics of nanoscale
  junctions from first principles}.
\newblock \emph{\bibinfo{journal}{Phys. Rev. B}} \textbf{\bibinfo{volume}{86}},
  \bibinfo{pages}{155411} (\bibinfo{year}{2012}).

\bibitem{EspositoRMP09}
\bibinfo{author}{Esposito, M.}, \bibinfo{author}{Harbola, U.} \&
  \bibinfo{author}{Mukamel, S.}
\newblock \bibinfo{title}{Nonequilibrium fluctuations, fluctuation theorems,
  and counting statistics in quantum systems}.
\newblock \emph{\bibinfo{journal}{Rev. Mod. Phys.}}
  \textbf{\bibinfo{volume}{81}}, \bibinfo{pages}{1665--1702}
  (\bibinfo{year}{2009}).

\bibitem{SchonPRB94}
\bibinfo{author}{Schoeller, H.} \& \bibinfo{author}{Sch{\"{o}}n, G.}
\newblock \bibinfo{title}{Mesoscopic quantum transport: Resonant tunneling in
  the presence of a strong {C}oulomb interaction}.
\newblock \emph{\bibinfo{journal}{Phys. Rev. B}} \textbf{\bibinfo{volume}{50}},
  \bibinfo{pages}{18436--18452} (\bibinfo{year}{1994}).

\bibitem{SchonPRL96}
\bibinfo{author}{K\"onig, J.}, \bibinfo{author}{Schoeller, H.} \&
  \bibinfo{author}{Sch\"on, G.}
\newblock \bibinfo{title}{Zero-bias anomalies and boson-assisted tunneling
  through quantum dots}.
\newblock \emph{\bibinfo{journal}{Phys. Rev. Lett.}}
  \textbf{\bibinfo{volume}{76}}, \bibinfo{pages}{1715--1718}
  (\bibinfo{year}{1996}).

\bibitem{SchoellerSchonPRB96}
\bibinfo{author}{K\"onig, J.}, \bibinfo{author}{Schmid, J.},
  \bibinfo{author}{Schoeller, H.} \& \bibinfo{author}{Sch\"on, G.}
\newblock \bibinfo{title}{Resonant tunneling through ultrasmall quantum dots:
  Zero-bias anomalies, magnetic-field dependence, and boson-assisted
  transport}.
\newblock \emph{\bibinfo{journal}{Phys. Rev. B}} \textbf{\bibinfo{volume}{54}},
  \bibinfo{pages}{16820--16837} (\bibinfo{year}{1996}).

\bibitem{PedersenWackerPRB05}
\bibinfo{author}{Pedersen, J.~N.} \& \bibinfo{author}{Wacker, A.}
\newblock \bibinfo{title}{Tunneling through nanosystems: Combining broadening
  with many-particle states}.
\newblock \emph{\bibinfo{journal}{Phys. Rev. B}} \textbf{\bibinfo{volume}{72}},
  \bibinfo{pages}{195330} (\bibinfo{year}{2005}).

\bibitem{WackerPRB11}
\bibinfo{author}{Karlstr\"om, O.}, \bibinfo{author}{Pedersen, J.~N.},
  \bibinfo{author}{Samuelsson, P.} \& \bibinfo{author}{Wacker, A.}
\newblock \bibinfo{title}{Canyon of current suppression in an interacting
  two-level quantum dot}.
\newblock \emph{\bibinfo{journal}{Phys. Rev. B}} \textbf{\bibinfo{volume}{83}},
  \bibinfo{pages}{205412} (\bibinfo{year}{2011}).

\bibitem{GrifoniEPJB13}
\bibinfo{author}{{Kern, Johannes}} \& \bibinfo{author}{{Grifoni, Milena}}.
\newblock \bibinfo{title}{Transport across an anderson quantum dot in the
  intermediate coupling regime}.
\newblock \emph{\bibinfo{journal}{Eur. Phys. J. B}}
  \textbf{\bibinfo{volume}{86}}, \bibinfo{pages}{384} (\bibinfo{year}{2013}).

\bibitem{GrifoniPRB15}
\bibinfo{author}{Dirnaichner, A.} \emph{et~al.}
\newblock \bibinfo{title}{Transport across a carbon nanotube quantum dot
  contacted with ferromagnetic leads: Experiment and nonperturbative modeling}.
\newblock \emph{\bibinfo{journal}{Phys. Rev. B}} \textbf{\bibinfo{volume}{91}},
  \bibinfo{pages}{195402} (\bibinfo{year}{2015}).

\bibitem{SchonPRB03}
\bibinfo{author}{Thielmann, A.}, \bibinfo{author}{Hettler, M.~H.},
  \bibinfo{author}{K\"onig, J.} \& \bibinfo{author}{Sch\"on, G.}
\newblock \bibinfo{title}{Shot noise in tunneling transport through molecules
  and quantum dots}.
\newblock \emph{\bibinfo{journal}{Phys. Rev. B}} \textbf{\bibinfo{volume}{68}},
  \bibinfo{pages}{115105} (\bibinfo{year}{2003}).

\bibitem{ChenOchoaMGJCP17}
\bibinfo{author}{Chen, F.}, \bibinfo{author}{Ochoa, M.~A.} \&
  \bibinfo{author}{Galperin, M.}
\newblock \bibinfo{title}{Nonequilibrium diagrammatic technique for {H}ubbard
  {G}reen functions}.
\newblock \emph{\bibinfo{journal}{J. Chem. Phys.}}
  \textbf{\bibinfo{volume}{146}}, \bibinfo{pages}{092301}
  (\bibinfo{year}{2017}).

\bibitem{SandalovIJQC03}
\bibinfo{author}{Sandalov, I.}, \bibinfo{author}{Johansson, B.} \&
  \bibinfo{author}{Eriksson, O.}
\newblock \bibinfo{title}{Theory of strongly correlated electron systems:
  Hubbard-anderson models from an exact hamiltonian, and perturbation theory
  near the atomic limit within a nonorthogonal basis set}.
\newblock \emph{\bibinfo{journal}{Int. J. Quant. Chem.}}
  \textbf{\bibinfo{volume}{94}}, \bibinfo{pages}{113--143}
  (\bibinfo{year}{2003}).

\bibitem{Fransson_2010}
\bibinfo{author}{Fransson, J.}
\newblock \emph{\bibinfo{title}{Non-Equilibrium Nano-Physics. A Many-Body
  Approach.}} (\bibinfo{publisher}{Springer}, \bibinfo{year}{2010}).

\bibitem{EspMGPRB15}
\bibinfo{author}{Esposito, M.}, \bibinfo{author}{Ochoa, M.~A.} \&
  \bibinfo{author}{Galperin, M.}
\newblock \bibinfo{title}{Efficiency fluctuations in quantum thermoelectric
  devices}.
\newblock \emph{\bibinfo{journal}{Phys. Rev. B}} \textbf{\bibinfo{volume}{91}},
  \bibinfo{pages}{115417} (\bibinfo{year}{2015}).

\bibitem{EspositoMGJPCC10}
\bibinfo{author}{Esposito, M.} \& \bibinfo{author}{Galperin, M.}
\newblock \bibinfo{title}{Self-consistent quantum master equation approach to
  molecular transport}.
\newblock \emph{\bibinfo{journal}{J. Phys. Chem. C}}
  \textbf{\bibinfo{volume}{114}}, \bibinfo{pages}{20362--20369}
  (\bibinfo{year}{2010}).

\bibitem{WernerRMP14}
\bibinfo{author}{Aoki, H.} \emph{et~al.}
\newblock \bibinfo{title}{Nonequilibrium dynamical mean-field theory and its
  applications}.
\newblock \emph{\bibinfo{journal}{Rev. Mod. Phys.}}
  \textbf{\bibinfo{volume}{86}}, \bibinfo{pages}{779--837}
  (\bibinfo{year}{2014}).

\bibitem{StefanucciVanLeeuwen_2013}
\bibinfo{author}{Stefanucci, G.} \& \bibinfo{author}{van Leeuwen, R.}
\newblock \emph{\bibinfo{title}{Nonequilibrium Many-Body Theory of Quantum
  Systems. A Modern Introduction.}} (\bibinfo{publisher}{Cambridge University
  Press}, \bibinfo{year}{2013}).

\bibitem{TalRuitenbeekPRL08}
\bibinfo{author}{Kiguchi, M.} \emph{et~al.}
\newblock \bibinfo{title}{Highly conductive molecular junctions based on direct
  binding of benzene to platinum electrodes}.
\newblock \emph{\bibinfo{journal}{Phys. Rev. Lett.}}
  \textbf{\bibinfo{volume}{101}}, \bibinfo{pages}{046801}
  (\bibinfo{year}{2008}).

\bibitem{WhiteGalperinPCCP12}
\bibinfo{author}{White, A.~J.} \& \bibinfo{author}{Galperin, M.}
\newblock \bibinfo{title}{Inelastic transport: a pseudoparticle approach}.
\newblock \emph{\bibinfo{journal}{Phys. Chem. Chem. Phys.}}
  \textbf{\bibinfo{volume}{14}}, \bibinfo{pages}{13809--13819}
  (\bibinfo{year}{2012}).

\end{thebibliography}

\section*{Acknowledgements}
This material is based upon work supported by the National Science Foundation under CHE - 1565939,
and JSPS KAKENHI (Grant No. 15J03915, 15H02025, 16K21623). 
Part of the numerical calculations were performed using HOKUSAI system at RIKEN.

\section*{Author contributions statement}
K.M. prepared diagrammatic expansion within PP- and Hubbard-NEGF, 
and performed noise spectrum simulations within PP-NEGF.
F.C. participated in developing the Hubbard NEGF. 
M.G. performed simulations within the Hubbard NEGF, and prepared the manuscript. 
All authors reviewed the manuscript. 

\section*{Additional information}
\textbf{Competing financial interests} The authors declare no competing financial interests. 

\end{document}